\documentclass{IEEEtran}
\usepackage[utf8]{inputenc}
\usepackage[T1]{fontenc}
\usepackage{graphicx}
\usepackage{grffile}
\usepackage{longtable}
\usepackage{wrapfig}
\usepackage{rotating}
\usepackage[normalem]{ulem}
\usepackage{amsmath}
\usepackage{textcomp}
\usepackage{amssymb}
\usepackage{xspace}
\usepackage{capt-of}
\usepackage{hyperref}
\usepackage{lastpage}
\usepackage{todonotes}
\usepackage{multicol}
\usepackage{tcolorbox}
\usepackage{pifont}
\usepackage{enumitem}
\newcommand\descmargin{12pt}

\author{
  \IEEEauthorblockN{Pierre \textsc{Alliez}},
  \IEEEauthorblockA{Universit\'e C\^ote d'Azur, Inria, France, \url{mailto:pierre.alliez@inria.fr}}\\
  \and
  \IEEEauthorblockN{Roberto \textsc{Di Cosmo}},
  \IEEEauthorblockA{Inria, Software Heritage, University of Paris, France, \url{mailto:roberto@dicosmo.org}}\\
  \and
  \IEEEauthorblockN{Benjamin \textsc{Guedj}},
  \IEEEauthorblockA{Inria, France and University College London, United Kingdom, \url{mailto:benjamin.guedj@inria.fr}}\\
  \and
  \IEEEauthorblockN{Alain \textsc{Girault}},
  \IEEEauthorblockA{Univ.\ Grenoble Alpes, Inria, CNRS, Grenoble INP, LIG, 38000 Grenoble, France, \url{mailto:alain.girault@inria.fr}}\\
  \and
  \IEEEauthorblockN{Mohand-Saïd \textsc{Hacid}},
  \IEEEauthorblockA{Univ. Lyon, University Claude Bernard Lyon 1, LIRIS, Lyon France, \url{mailto:mohand-said.hacid@univ-lyon1.fr}}\\
  \and
  \IEEEauthorblockN{Arnaud \textsc{Legrand}},
  \IEEEauthorblockA{Univ.\ Grenoble Alpes, CNRS, Inria, Grenoble INP, LIG, 38000 Grenoble, France, \url{mailto:arnaud.legrand@inria.fr}}\\
  \and
  \IEEEauthorblockN{Nicolas \textsc{Rougier}},
  \IEEEauthorblockA{Univ.\ Bordeaux, Inria, CNRS, IMN, Labri, Bordeaux, France, \url{mailto:nicolas.rougier@inria.fr}}
}

\date{\today}
\title{Attributing and Referencing (Research) Software: Best Practices and Outlook from Inria}
\hypersetup{
 pdfauthor={GT author list},
 pdftitle={Attributing and Referencing (Research) Software: Best Practices and Outlook from Inria},
 pdfkeywords={Software, Citation, Reference, Best Practices, Inria},
 pdfsubject={},
 pdfcreator={Emacs 26.1 (Org mode 9.1.14)}, 
 pdflang={English}}
\begin{document}

\newcommand{\fnurl}[2]{#1: \href{#2}{#2}}
\let\urlOld=\url
\def\url#1{\href{#1}{\textsf{#1}}}

\newcounter{proposal}
\newenvironment{proposal}[1]{\refstepcounter{proposal} 
  \begin{tcolorbox}[colback=red!5!white,colframe=red!75!black,title=\textbf{Recommendation \#\arabic{proposal}: #1}]}{
  \end{tcolorbox}
}

\maketitle

\begin{abstract}
  Software is a fundamental pillar of modern scientific research,
  across all fields and disciplines. However, there is a lack of
  adequate means to cite and reference software due to the complexity
  of the problem in terms of authorship, roles and credits. This
  complexity is further increased when it is considered over the
  lifetime of a software that can span up to several decades.
  Building upon the internal experience of Inria, the French research
  institute for digital sciences, we provide in this paper a
  contribution to the ongoing efforts in order to develop proper
  guidelines and recommendations for software citation and
  reference. Namely, we recommend: (1)~a richer taxonomy for software
  contributions with a qualitative scale; (2)~to put humans at the
  heart of the evaluation; and (3)~to distinguish citation from
  reference.

\end{abstract}

\section{Introduction}
\label{sec:intro}

Software is a fundamental pillar of modern scientific research, across
all fields and disciplines,
and the actual \emph{knowledge} embedded in software is contained in
software \emph{source code} which is, as written in the GPL license,
``the preferred form [of a program] for making modifications to it [as
  a developer]''
and ``provides a view into the mind of the
designer''~\cite{shustek06}. With the rise of Free/Open Source
Software, which requires and fosters source code accessibility, access
has been provided to an enormous amount of software source code that
can be massively reused. Similar principles are now permeating the
Open Science movement,
in particular after the attention drawn to it by the crisis in
scientific reproducibility~\cite{Stodden-reprod-2012,Hinsen2013}. All
this has recently motivated the need of properly referencing and
crediting software in scholarly
works~\cite{HowisonBullard2016,SoftwareCitationPrinciples-2016,swhipres2018}.

In this context, we provide a contribution to the ongoing efforts to
develop proper guidelines and recommendations, building upon the
internal experience of Inria, the French research institute for
digital sciences (\url{http://www.inria.fr}). Born in 1967, more 50
years ago, Inria has grown to directly employ 2,400 people, and its
190 project-teams involve more than 3,000 scientists working towards
meeting the challenges of computer science and applied mathematics,
often at the interface with other disciplines. Software lies at the
very heart of the Institute's activity, and it is present in all its
diversity, ranging from very long term large systems (e.g., the award
winning Coq proof assistant, to the CompCert certified compiler,
through the CGAL Computational Geometry Algorithms Library to name
only a few of most the well-known ones), to medium sized projects and
small but sophisticated codes implementing advanced algorithms.

Inria has always considered software as a first class noble product of
research, as an instrument for research itself, and as an important
contribution in the career of researchers. As such, whenever a team is
evaluated or a researcher applies for a position or a promotion, a
concise and precise self-assessment notice must be provided for each
software developed in the team or by the applicant, so that it can be
assessed in a systematic and relevant way.

With the emerging awareness of the importance of making research
openly accessible and reproducible, Inria has stepped up its
engagement for software: (i)~it has been working for years on
reproducible research, and is running a MOOC on this subject; (ii)~it
has been at the origin of the Software Heritage initiative, which is
building a universal archive of source code~\cite{swhcacm2018}; and
(iii) it has experimented a novel process for research software
deposit and citation by connecting the French national open access
publication portal, HAL (\url{hal.archives-ouvertes.fr}), to Software
Heritage~\cite{swhcacm2018}.

Nevertheless, citing and referencing software is a very complex task, for
several reasons. First, software authorship is extremely varied,
involving many roles: software architect, coder, debugger, tester,
team manager, and so on. Second, software itself is a complex object:
the lifespan can range from a few months to decades, the size can
range to a few dozens of lines of code to several millions, and it can
be stand-alone or rely on multiple external packages. And finally,
sometimes one may want to reference a particular version of a given
software (this is crucial for reproducible research), while at other
times one may want to cite the software as a whole.

In this article, we report on the practices, processes, and vision,
both in place and under consideration at Inria, to address the
challenges of referencing and accessing software source code, and
properly crediting the people involved in their development,
maintenance, and dissemination.

The article is structured as follows: Section~\ref{sec:survey} briefly
surveys previous work. Section~\ref{sec:complex} describes the
inherent complexity of software, which is the main reason why the
topic studied in this paper is challenging. Section~\ref{sec:inria}
presents the key internal processes that Inria has established over
the last decades to track the hundreds of software projects to which
the institute contributes, and the criteria and taxonomies they
use. Section~\ref{sec:lessons} draws the main lessons that have been
learned from this long term experience. In particular, we state three
recommendations to contribute to a better handling of research
software worldwide. Finally, Section~\ref{sec:conclusion} concludes by
providing a set of recommendations for the future.

\def\alvinbreak{\discretionary{}{}{}}

\section{Survey of previous work}
\label{sec:survey}

The astrophysics community is one of the oldest ones having attempted
to systematically describe the software developed and use in their
research work. The Astrophysics Source Code Library was started in
1999. Over the years it has established a curation process that
enables the production of quality metadata for research
software. These metadata can be used for citation purposes, and they
are widely used in the astrophysics field~\cite{ASCL}. Around 2010,
interest in software arose in a variety of domains: a few computer
science conferences started an \emph{artefact evaluation} process
(see \href{https://www.artifact-eval.org/}{www.artifact-eval.org})
which has spread to many top venues in computer science. This led to
the badging system that ACM promotes for articles presenting or using
research software~\cite{AcmBadges} and to the cloud-based software
hosting solution used and put forward by IEEE and Taylor \& Francis
for their journals (Code Ocean). The need to take research software
into account, making it available, referenceable, and citable, became
apparent in many research
communities~\cite{Borgman2012,Stodden-reprod-2012,Hinsen2013,gil2016},
and the limitation of the informal practices currently in use quickly
surfaced~\cite{HowisonBullard2016,Collberg2016,Hwang2017}. An
important effort to bring together these many different experiences,
and to build a coherent point of view has been made by the FORCE11
Software Citation Working Group in 2016, which led to state a concise
set of \emph{software citation
  principles}~\cite{SoftwareCitationPrinciples-2016}. In a nutshell,
this document recognizes the importance of software, credit and
attribution, persistence and accessibility, and provides several
recommendations based on use-cases that illustrate the different
situations where one wants to cite a piece of software.

We do acknowledge these valuable efforts, which have contributed to
raise the awareness about the importance of research software in the
scholarly world.

Nonetheless, we consider that a lot more work is needed before we can
consider this problem settled: the actual \emph{recommendations} that
can be found on how to make software citable and referenceable, and
how to give credit to its authors, fall quite short of what is needed
for an object as complex as software. For example, in most of the
guidelines we have seen, making software \emph{referenceable} for
reproducibility (where the precise version of the software needs to be
indicated), or \emph{citable} for credit (to authors or institutions),
seems to boil down to simply finding a way to attach a DOI
to it, typically by depositing a copy of the source code in
repositories like Zenodo or Figshare.

This simple approach, inspired by common practices for research data,
is not appropriate for software.

When our goal is giving credit to authors, attaching an identifier to
metadata is the easy part, and any system of digital identifiers, be
it DOI, Ark or Handles, will do. The difficulty lies in getting
quality metadata, and in particular in determining \emph{who should
get credit}, \emph{for what kind of contribution}, and \emph{who has
authority to make these decisions}. The heated debate spawned by
recent experiments that tried to automatically compute the list of
authors out of commit logs in version control
systems~\cite{TBrownRevisited} clearly shows how challenging this can
be.

As we will see in Section~\ref{ssec:reprod}, when looking for
reproducibility, it is necessary to precisely identify not only the
main software but also its whole environment and to make it available
in an open and perennial way. In this context, we need verifiable build
methods and intrinsic identifiers that do not depend on resolvers that
can be abused or compromised (see Wiley using fake DOIs to trap web
crawlers \ldots\ and researchers as well), and DOIs are not designed
for this use case~\cite{swhipres2018}.

\begin{quote}
  To make progress in our effort to make research software better
  recognized, a first step is to acknowledge its complexity, and to
  take it fully into account in our recommendations.
\end{quote}

\def\alvinbreak{\discretionary{}{}{}}

\section{Complexity of the software landscape}
\label{sec:complex}

Software is very different from articles and data, with which we have
much greater familiarity and experience, as they have been produced
and used in the scholarly arena long before the first computer program
was ever written. In this section, we highlight some of the main
characteristics that render the task of assessing, referencing,
attributing and citing software a problem way more complex than what
it may appear at first sight.

Software development is a multifaceted and continuously evolving activity,
involving a broad spectrum of goals, actors, roles, organizations, practices
and time extents. Without pretending to be exhaustive, we detail here in bold
the most important aspects that need to be taken into account for assessing,
referencing, attributing or citing software.

\begin{description}[leftmargin=\descmargin]

\item[\textbf{Structure}]\mbox{}\\ A software project can be organized
  either as a \emph{monolithic program} (e.g., the Netlib BLAS
  libraries), or as a \emph{composite assembly of modules} (e.g., the
  Clang compiler). It can either be \emph{self-contained} or have many
  \emph{external dependencies}. For example, the Eigen C++ template
  library for linear algebra (\url{http://eigen.tuxfamily.org})
  aims for minimal dependencies while listing an ecosystem of
  unsupported modules.

\item[\textbf{Lifetime}]\mbox{}\\ A software can be produced during a
  single, short extent of time (referred to as \emph{one-shot
    contribution}), or over a long time span, possibly fragmented into
  several time intervals of activities. Some long running software
  projects extend over several decades. For example, the CGAL project
  (\url{https://www.cgal.org/project.html}) started in 1996 as a
  European consortium, became open source in 2004, and has provided
  more than 30 releases since then.

\item[\textbf{Community}]\mbox{}\\ A software can be the product of a
  single scholar, a well-identified team or a scattered team of
  scholars spanning a large scientific community that may be difficult
  to track precisely. The CGAL open source project lists more than 130
  contributors, distinguishing between the former and current
  developers, and acknowledging the reviewers and initial consortium
  members (\url{https://www.cgal.org/people.html}). In contrast, the
  Meshlab 3D mesh processing software
  (\url{https://en.wikipedia.org/wiki/MeshLab}) is authored by a
  single team from the CNR, Pisa.

\item[\textbf{Authorship}]\mbox{}\\ Software developer(s) writing the
  code are the most visible authors of a software program, but they
  are not, and by far, the only ones. A variety of activities are
  involved in the creation of software, ranging from stating the
  high-level specifications, to testing and bug fixing, through
  designing the software architecture, making technical choices,
  running use cases, implementing a demonstrator, drafting the
  documentation, deploying onto several platforms, and building a
  community of users. In these contexts the roles of a single
  contributor can be plural, with contributions spanning variable time
  extents. Authorship is even more complicated when developers resort
  to pseudonimity, i.e., disguised identity in order to not disclose
  their legal identities. For all these reasons, evaluating the real
  contributions to a significant piece of software is a very difficult
  problem: in our experience at Inria, automated tools may help in
  this task, but are by far insufficient, and it is essential to have
  humans in the loop.

\item[\textbf{Authority}]\mbox{}\\ Beyond good practices, most quality
  or certified software development projects define management
  processes and authority rules. Authorities are entitled to make
  decisions, give orders, control processes, enforce rules, and
  report. They can be institutions, organizations, communities, or
  sometimes a single person (e.g., Guido van Rossum for Python). Some
  projects set up an editorial board, similar in spirit to scientific
  journals, with reviewers, managers and well-defined procedures (See
  CGAL's Open Source Project Rules and Procedures at
  \url{https://www.cgal.org/project\_rules.html}). Each new
  contribution must be submitted for review and approval before being
  integrated. Some decisions can be taken top-down while others are
  bottom-up. In some cases, a shared governance is implemented. This
  organization can be somehow compared to the Linux kernel development
  organization where Linus Torvalds integrates contributions but
  delegates the responsibility of software quality evaluation to a few
  trusted colleagues. Another important aspect is the traceability of
  who did what during the software project. In its simplest form, the
  number of lines or code or commit logs are used for tracing
  contributions and changes, but more advanced means such as
  repository mining-based metrics~\cite{lima-assess-2015}, bug-related
  metrics, or peer evaluation are common.

\item[\textbf{Levels of description}]\mbox{}\\ Another dimension that
  adds to the complexity is the variety of levels at which a software
  project can be described, either for citation or for reference.

  \textbf{Exact status of the source code.} For the purpose of exact
  reproducibility, one must be able to \emph{reference} any precise
  point in the development history of a software project, even if it
  is not labeled as a release; in this case, cryptographic identifiers
  like those used in distributed version control systems, and now
  generalized in Software Heritage~\cite{swhipres2018}, are
  \emph{necessary}. For instance, the sentence \emph{``you can find at
    \href{https://archive.softwareheritage.org/swh:1:cnt:cdf19c4487c43c76f3612557d4dc61f9131790a4;lines=146-187/}{swh:1:cnt:cdf19c4487c43c76f3612557d4dc61f9131790a4;\alvinbreak{}lines=146-187}
    of
    \href{https://archive.softwareheritage.org/swh:1:snp:c9c31ee9a3c631472cc8817886aaa0d3784a3782;origin=https://github.com/rdicosmo/parmap/}{swh:1:snp:c9c31ee9a3c631472cc881788\alvinbreak{}6aaa0d3784a3782;origin=https://github.com/rdicosmo/\alvinbreak{}parmap/}
    the exact core mapping algorithm used in this article''} makes two
  distinct references. The former one points to the lines of a source
  file while the later one points to the software context in which
  this file is used.

  \textbf{(Major) release.} When a much coarser granularity is
  sufficient, one can designate a particular (major) release of the
  project. For instance: \emph{``This functionality is available in
    OCaml version 4''} or \emph{``from CGAL version~3''}.

  \textbf{Project.} Sometimes one needs to \emph{cite a software
    project} at the highest level; a typical example is a researcher,
  a team or an institution reporting the software projects it develops
  or contributes to. In this case, one must list only \emph{the
    project as a whole}, and not all its different versions. For
  instance: \emph{``Inria has created OCaml and Scikit-Learn''}.

\end{description}

\section{Four processes for four different needs}
\label{sec:inria}

There are four main reasons why the research software produced at
Inria is carefully referenced and evaluated: (i)~managing the career
of individual researchers and research engineers, (ii)~assessing the
technology transfer, (iii)~visibility and impact of a research
team, and (iv)~promote reproducible research practices.
We detail next these four topics, and the information collected, 
to cater to these different needs.

\subsection{Career management}
\label{ssec:career}

Software development is a research output taken into account in the
evolution of the \emph{career of individual researchers and research
  engineers}. Measuring the impact of a software provides a means to
measure the scope and magnitude of contributions of research results,
when they are carefully translated into usable software. Evaluating
the maturity and breadth of software is also essential to guide
further developments and resource allocation.

Inria has an internal evaluation body, the Evaluation Committee (EC),
the role of which includes evaluating both individual researchers when
they apply for various positions (typically ranging from junior
researcher to leading roles such as senior researcher or research
director), and organizing the evaluations of whole research teams,
which take place every 4 years. In both cases, evaluating a given
software revolves around three items: (i)~the software itself, which
can be downloaded and tested; (ii)~precise self-assessment criteria
filled-in by the developers themselves; and (iii)~a factual and
high-level description of the software, including the programming
language(s) used along with the number of lines of code, the number of
man-months of development effort, and the web site from where the
software and any other relevant material (a user manual, demos,
research papers, ...) can be downloaded.

Among these three items, the self-assessment criteria play a crucial
role because they provide key information on the software, how it was
developed, and what role each developer played. Version~1 of these
``Criteria for Software Self-Assessment'' dates from August
2011~\cite{sw-criteria-inria}. They are also used by the Institute for
Information Sciences and Technologies (INS2I) of The French National
Centre for Scientific Research (CNRS). It comprises two lists of
criteria using a \emph{qualitative} scale. 
The first list characterizes the software itself:
\newcommand\criteria[1]{\textbf{\textsf{#1}}\xspace}

\begin{description}[leftmargin=\descmargin]
\item[\criteria{Audience}.] Ranging from \criteria{A1} (personal
  prototype) to \criteria{A5} (usable by a wide public).

\item[\criteria{Software Originality}.] Ranging from
  \criteria{SO1} (none) to \criteria{SO4} (original software implementing
  a fair number of original ideas).

\item[\criteria{Software Maturity}.] Ranging from \criteria{SM1}
  (demos work, rest not guaranteed) to \criteria{SM5} (high-assurance
  software, certified by an evaluation agency or formally verified).

\item[\criteria{Evolution and Maintenance}.] Ranging from
  \criteria{EM1} (no future plans) to \criteria{EM4} (well-defined and
  implemented plan for future maintenance and evolution, including an
  organized users group).

\item[\criteria{Software Distribution and Licensing}] Ranging
  from \criteria{SDL1} (none) to \criteria{SDL5} (external packaging and
  distribution either as part of e.g., a Linux distribution, or
  packaged within a commercially distributed product).
\end{description}

As an example, the OCaml compiler is assessed as: Audience
\criteria{A5}, Software Originality \criteria{SO3}, Software Maturity
\criteria{SM4}, Evolution and Maintenance \criteria{EM4}, Software
Distribution and Licensing \criteria{SDL5}.

The second list characterizes the contribution of the developers and
comprises the following criteria: \criteria{Design and Architecture
  (DA)}, \criteria{Coding and Debugging (CD)}, \criteria{Maintenance
  and Support (MS)}, and \criteria{Team/Project Management
  (TPM)}. Each contribution ranges from \criteria{1} (not involved) to
\criteria{4} (main contributor).  As an example, the personal
contribution of one of OCaml's main developer might be: Design and
Architecture \criteria{DA3}, Coding and Debugging \criteria{CD4},
Maintenance and Support \criteria{MS3}, Team/Project Management
\criteria{TPM4}.

Overall, these self-assessment criteria have been in used at Inria for
several years now. The feedback from both jury members (for individual
researchers) and international evaluators (for research teams) is that
they are extremely useful, despite their coarse granularity and being
based on self-statement. All praise the relevance of the criteria and
the fact that they provide a mean to assess the scope and magnitude of
contributions to a given software, much more accurately.

\begin{figure*}[t]
  \includegraphics[width=\textwidth]{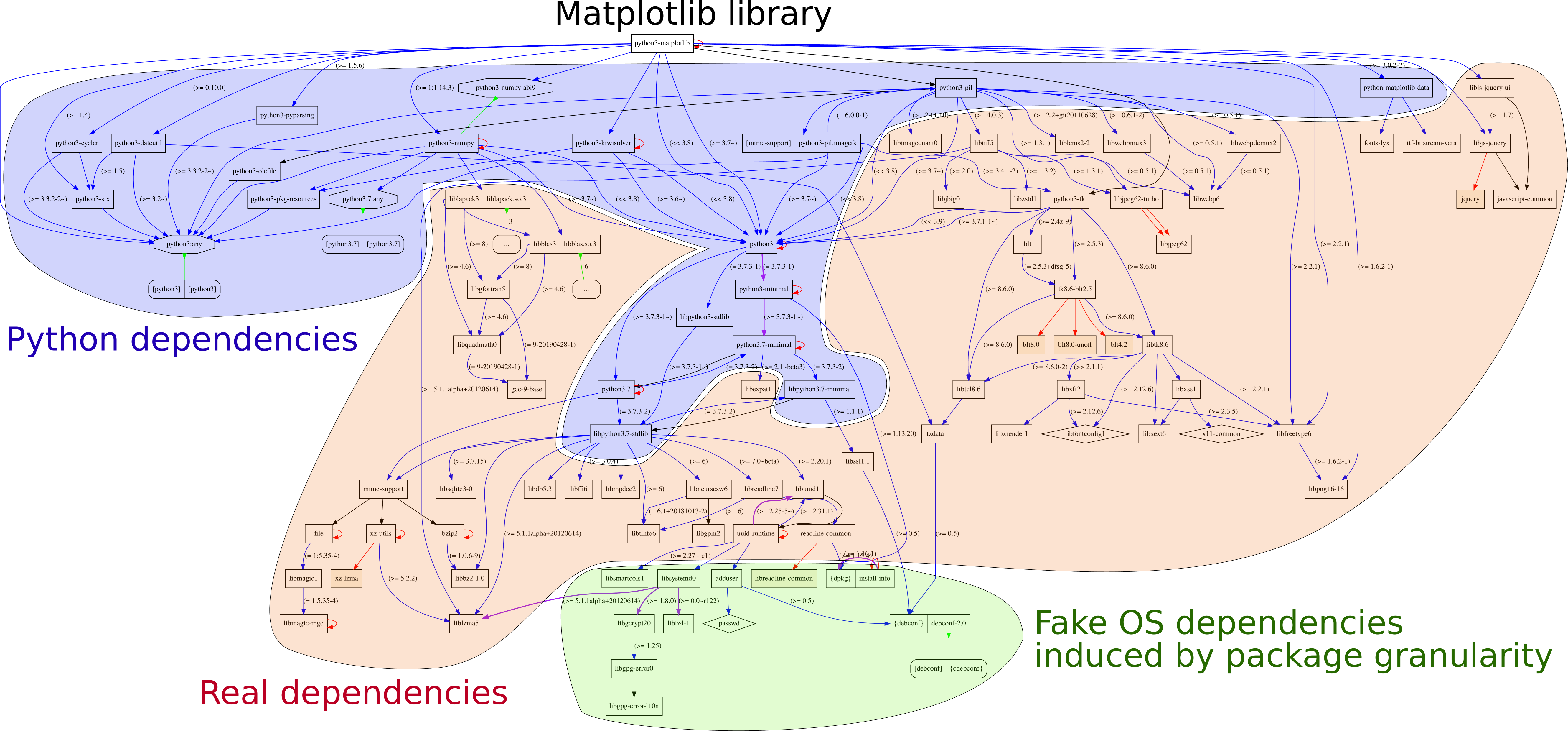}
  \caption{Example of the complexity in direct and indirect
    dependencies for a specific python package (matplotlib). Boxes
    represent actual packages (libraries that need to be installed on
    the system), arrows indicates dependencies to other packages,
    labels indicates the minimal/maximal version number. In blue the
    Python dependencies, in red the ``true'' system dependencies
    incurred by python (e.g., the \texttt{libc} or
    \texttt{libjpeg62}), in green some ``fake'' dependencies incurred by the
    package management system but which are very likely not used by python
    (e.g., \texttt{adduser} or \texttt{dpkg}). \label{fig:deptree}}
            \end{figure*}

\subsection{Technology transfer}
\label{ssec:transfer}

Information about authorship and intellectual property is a key asset
when \emph{technology transfer} takes place, either in industrial
contracts or for the creation of start-ups. Besides, technology
transfer is at the heart of Inria's strategy to increase its societal
and economical impact. However, in the particular case of software,
technology transfer raises a number of difficulties. Most of the time,
transferring a software to industry starts by sending a copy of the
software to a French registration agency named \emph{Agence pour la
  Protection des Programmes} (APP:
\url{https://www.app.asso.fr}). When doing so, a dedicated form has to
be filled that requires to specify \emph{all the contributors} of the
software, and for each of them the \emph{percentage} of her/his
contribution.

When the software is old (typically more than 10 years old), this
involves carrying on some archaeology to retrieve the contribution of
the first developers (some of whom may have left Inria, or may have
not been Inria employees at all). A dedicated technology transfer team
interacts with the researchers in this process, taking into account
all the different contributions to software development. In
particular, they use a taxonomy of roles that includes the following:

\begin{description}[leftmargin=\descmargin]

\item[\textbf{Coding}.]\mbox{}\\ This seems the most obvious part, but
  it is actually complex, as one cannot just count the number of lines
  of codes written, or the number of accepted pull requests. Sometimes
  a long code fragment may be a straightforward re-implementation of a
  very well known algorithm or data structure, involving no complexity
  or creativity at all, while at other times a few lines of code can
  embody a complex and revolutionary approach (e.g., speeding up
  massively the execution time). Often, a major contribution to a
  project is not adding code, but fixing code or removing portions of
  code by factoring the project and increasing its modularity and
  genericity.

\item[\textbf{Testing and debugging}.]\mbox{}\\ This is an essential
  role when developing software that is meant to be used more than
  once. This activity may require setting up a large database of
  relevant use cases and devising a rigorous testing protocol (e.g.,
  non-regression testing).

\item[\textbf{Algorithm design}.]\mbox{}\\ Inventing the underlying
  algorithm that forms the very basis of the software being
  transferred to industry is, of course, a key contribution.

\item[\textbf{Software architecture design}]\mbox{}\\ This is another
  important activity that does not necessarily show up in the code
  itself, but which is essential for maintenance, modularity,
  efficiency and evolution of the software. As Steve Jobs famously
  said while promoting Object Oriented Programming and the NeXT
  computer more than twenty-five years ago, \emph{``The line of code
    that has no bug and that costs nothing to maintain, is the line of
    code that you never wrote''}.

\item[\textbf{Documentation}.]\mbox{}\\ This activity is essential to
  ease (re)usability and to support long term maintenance and
  evolution. It ranges from internal technical documentation to
  drafting the user manual and tutorials.
\end{description}

The older and bigger the software, the more difficult this authorship
identification task is.

\subsection{Visibility and impact of a research team}
\label{ssec:visibility}

Software is a part of the scientific production that any research team
exposes. Software that are diffused to a large scholar audience or
commercialized to industrial users may become an important source of
inspiration for novel research challenges. Feedback from practitioners
or academic users is a precious source of knowledge for determining
the research problems with high potential practical impact. Software
can also be a key instrument for research, central to the daily
research activity of a team, and a main support for teaching and
education. It may also become a communication medium between young
researchers, e.g., Ph.D.\ students sharing their research topics and
experiments via a common set of software components.

Inria considers (research) software to be a valuable output of
research, and has always encouraged its research teams to advertise
the software project they contribute to: this can be on the public web
page of the team, or in its annual activity report. To simplify the
collection of the information concerning the software projects, an
internal database, called BIL (``Base d'Information des Logiciels'',
i.e.\ ``database of information on software''), has been in use for
several years. It allows research teams to deposit very detailed
meta-data describing the software projects they are involved in. The
BIL can then be used to generate automatically the list of software
descriptions for the team web page, for the activity report, and also
to prefill part of the forms used in the two processes described above
for individual career evaluation and for technology transfer, avoiding
the burden of typing in the same information over and over again.

\subsection{Reproducible Research}
\label{ssec:reprod}

Another important concern of Inria is supporting
\emph{reproducibility} of research results and the reproducibility
crisis takes a whole new dimension when software is involved. Scholars
are struggling to find ways to aggregate in a coherent
\emph{compendium} the data, the software, and the explanations of
their experiments. The focus is no longer on giving credit, but on
finding, rebuilding, and running the \emph{exact software referenced
  in a research article}. We identified at least three major issues:

First, the frequent lack of availability of the \emph{software source
  code}, and/or of precise references to the right version of it, is a
major issue~\cite{Collberg2016}. Solving this issue requires stable
and perennial source code archives and specialized
identifiers~\cite{swhipres2018}.

Second, characterizing and reproducing the \emph{full software
  environment} that is used in an experiment requires tracking a
potentially huge graph of dependencies (a small example is shown in
Figure~\ref{fig:deptree}). Specific tools to identify and express such
dependencies are needed. 
Finally, although the notion of research compendium is seducing, it 
should aggregate objects of very different nature (article, data,
software) for which specific archives and solutions may already
exist. To ease the deposit of such objects, we believe the compendium
should thus rather build on stable references to objects than try to
address all problems at once.

In recent years, various building blocks have emerged to address these
challenges and may lead to such a global approach and stable
\emph{references} to the software artifact themselves. Inria has
fostered and supported a few of them, that we briefly present here.

\begin{figure}[htbp]
  \centering
  \fbox{\includegraphics[width=.8\linewidth]{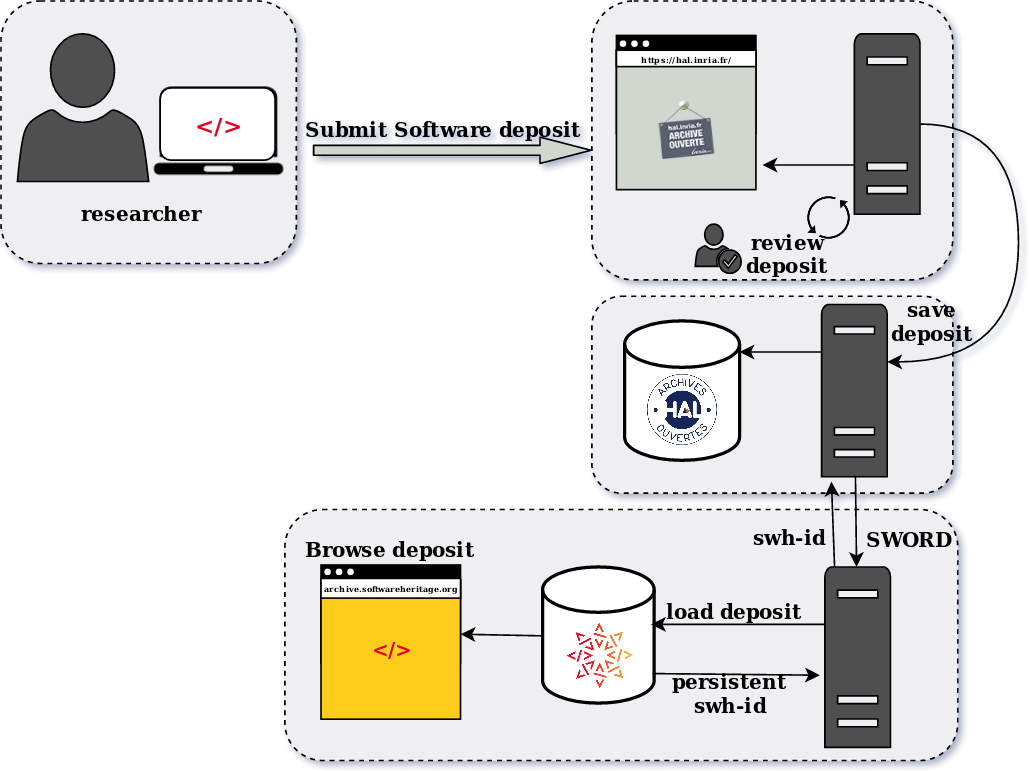}}
  \caption{Moderated software deposit in SWH via HAL.}
  \label{fig:swhdeposit}
\end{figure}

\begin{description}
\item[Software Heritage: a universal archive of source code.]
  Software Heritage (SWH) was started in 2015 to collect, preserve and
  share the source code of all software ever written, together with
  its full development history~\cite{swhcacm2018}. As of today, it has
  already collected almost 6 billions unique source code files coming
  from over 85 million software origins that are regularly
  harvested. The recently added ``save code now'' feature enables
  users to request proactively the addition of new software origins or
  to update them. Source code and its development history are stored
  in a \emph{universal data model} based on Merkle
  DAGs~\cite{swhipres2018}, providing \emph{persistent, intrinsic,
    unforgeable, and verifiable identifiers} for the more than 10
  billion objects it contains~\cite{swhipres2018}. Each intrinsic
  identifier is computed on the content and meta-data of the software
  itself, through cryptographic hashes, and is embedded into the
  software's persistent identifier.
  This \emph{universal archive} of \emph{all} software
  source code addresses the issue of \emph{preserving and referencing
    source code} for reproducibility.

\item[Reproducible builds.]\mbox{}\\
  In the early 2000's, the ground-breaking notion of \emph{functional
    package manager} was introduced by the Nix system~\cite{Nix2004},
  using cryptographic hashes to ensure that binaries are rebuilt and
  executed in the exact same software environment. Similar notions
  provide the foundation of the Guix toolchain, which has been
  developed over the last decade under the umbrella of the GNU
  project, with key contributions from Inria~\cite{Guix2015}. The
  essential property of these tools is that, given the same source
  files and the associated \emph{functional build recipes}, one can
  obtain as a result of the build process the very same binary files
  in the same environment.
  Very recently, Guix has been connected with SWH to ensure long term
  reproducibility: when the source code (currently downloaded from the
  upstream distribution sites) disappears from the designated
  location, Guix uses transparently the SWH intrinsic identifiers to
  fetch the archived copy from its archive. Functional build recipes
  are themselves a form of source code, and they too can be archived
  and given intrinsic identifiers, which will provide proper
  \emph{references} also for software environments.

\item[Curation of software deposit in HAL for SWH.]\mbox{}\\
  Over the past two years, Inria has fostered a collaboration between
  SWH and HAL, the French national open access archive, with the goal
  of providing an efficient process of research software
  deposit. Figure~\ref{fig:swhdeposit} provides a high level overview
  of this process: researchers submit software source code and
  meta-data to the HAL portal; these submissions are placed in a
  moderation loop where humans interact with the researchers to
  improve the quality of the meta-data and to avoid duplicates; once a
  submission is approved, it is sent to SWH via a generic deposit
  mechanism, based on the SWORD standard archive exchange protocol; it
  is then ingested in the SWH archive; finally, the unique intrinsic
  identifier needed for reproducibility is returned to the HAL portal,
  which displays it alongside the identifier for the
  meta-data. Detailed guidelines have been developed to help
  researchers
  and moderators
  get to a high quality deposit of
  their source code.

\end{description}

\def\alvinbreak{\discretionary{}{}{}}

\section{Lessons learned on crediting software}
\label{sec:lessons}

The processes described above have been established inside Inria and
refined over decades to answer the internal needs of the
institution. While their goal has not been to guide external processes
such as software citation, we strongly believe they provide a solid
basis to build a universal framework for software citation and
reference.

Here are a few important lessons we learned from all the above:
(research) software projects present a \emph{great degree of
  variability} along many axes; contributions to software can take
\emph{many forms and shapes}; and there are \emph{key contributions
  that must be recognised but do not show up in the code nor in the
  logs of the version control systems}. This has several main
consequences:

\begin{itemize}
\item the need of a \emph{rich metadata schema} to describe software
  projects;

\item the need of a \emph{rich taxonomy} for software contributions,
  that must not be flattened out on the simple role of software
  developer;

\item last but not least, while tools may help, a \emph{careful human
  process involving the research teams} is crucial to produce the
  qualified information and metadata that is needed for proper credit
  and attribution in the scholarly world.
\end{itemize}

We focus here mostly on the two latter issues, as the question of
metadata for software has already attracted significant attention,
with the Codemeta initiative providing a good vehicle for
standardisation, and for incorporating the new entities that may be
needed~\cite{CodeMeta}.

\subsection{Taxonomy of contributor roles: a proposal}
\label{ssec:taxonomy}

The need to recognise different levels and forms of contributions is
not new in academia: in Computer Science and Mathematics we are quite
used to separate, for example, the persons that are named as authors,
and those that are only mentioned in the acknowledgements.

In the specific case of software projects, the Software Credit
Ontology
\url{https://dagstuhleas.github.io/\alvinbreak{}SoftwareCreditRoles/doc/index-en.html}
proposes a total of 23 roles, among which 13 are directly concerned
with an actual contribution to a software project, under the
contributor category: code reviewer; community contributor; designer;
developer; documenter; idea contributor; infrastructure supporter;
issue reporter; marketing and sales; model driven software engineering
expert; packager; requirement elicitator; systems and network
engineer. As we can see, this ontology in more focused on the business
aspect of software projects (see for instance the marketing and sales
role) than on technology aspect (for instance, the developer role is
further refined into bug fixer, core developer, and maintainer). The
taxonomy we propose in the recommendation below is a refinement and
combination of the taxonomies presented in Section~\ref{ssec:career}
and~\ref{ssec:transfer}.

\begin{proposal}{A richer taxonomy for software contributions, with a
  qualitative scale}
  \label{rec:taxo}
  Giving credit to contributors of a software project is very similar
  to giving credit to contributors to research articles. We thus need
  a rich taxonomy. In the previous sections we discussed two
  taxonomies, developed and used in two different contexts inside
  Inria: despite minor differences (for example, maintenance and user
  support are not taken into account for technology transfer), one can
  extract rather easily the following taxonomy of contributor roles
  that covers all the use case seen, and that may be extended in the
  future:
  \begin{tcolorbox}
                                                                \bf
    \newcommand\bul{\ensuremath{\bullet}\xspace}
    \vspace{-1em}\hspace{-1.8em}\begin{tabular}{p{.26\linewidth}@{ }p{.35\linewidth}p{.42\linewidth}}
    \bul Design   &  \bul Debugging     &  \bul Maintenance    \\
    \bul Coding   &  \bul Architecture  &  \bul Documentation  \\
    \bul Testing  &  \bul Support       &  \bul Management
    \end{tabular}\vspace{-1em}
  \end{tcolorbox}

  But this is only part of the story: in both of the internal Inria
  processes we described, contributions are not just \emph{classified}
  in different roles, they are also \emph{quantified}, either at a
  coarse grain (from 1 to 5 for career evaluation), or at a very fine
  grain (percentages are used for technology transfer, where a
  financial return needs to be precisely redistributed). We thus recommend
  using a coarse grain qualitative scale as it is easy to implement
  and proves to be very helpful whenever technology transfer occurs.
\end{proposal}

Other disciplines too have pushed efforts to create a richer taxonomy
of contributions for research articles, with the CRediT
system~\cite{creditpaper} detailing 14 different possible roles, one
of which is \emph{software}: the key idea is that each person listed
as an author needs to specify one or more of the 14 roles.

\subsection{The importance of the human in the loop}
\label{ssec:humanintheloop}

This quantification is essential, in particular considering that an
academic credit system will be inevitably built on top of software
citations, which brings us to our next key point: the importance of
having humans in the loop, which has already been clearly advocated in
a different context by the team behind the Astronomic Source Code
Library~\cite{ASCL}.

As we have already noted, many of the contributor roles identified
above are not reflected in the code. In order to assess these roles,
in kind and quantity, it is necessary to interact with the team that
has created and evolved the software: this is what the technology
transfer service at Inria routinely does.

What about the activities that are tightly related to the software
source code itself, like coding, testing, and debugging? Here it is
very tempting to try to use automated tools to determine the role of a
contributor, and the importance of each contribution. There are indeed
a wealth of different developer scoring algorithms that target GitHub
contributors (see for example \url{http://git-awards.com/},
\url{https://github.com/msparks/git-score} and GitHub's own scoring
using the number of commits, deletions, or additions). Unfortunately
these measures are far from robust: refactoring (that may be just
renaming or moving file around or even changing tabs in spaces!) can
lead to huge score increases, while the actual developer contribution
is marginal. And even if one could rule out irrelevant code changes,
our experience at Inria is that the importance and quality of a
contribution cannot be assessed by counting the number of lines of
code that have been added (see our description of the coding role in
Section~\ref{ssec:transfer}). This is particularly the case for
\emph{research} software that involves significant innovations.

\begin{proposal}{Putting human at the heart of the evaluation}
  \label{rec:human}
  As a bottomline, we strongly suggest to refrain, for \emph{research
    software}, from trying to generate software citation and credit
  metadata, and in particular the list of (main) authors, using
  automated tools: we need instead quality information in the scholarly world,
  and currently this can only be achieved with qualified human
  intervention. We strongly encourage the authors of research software
  to provide such qualitative information, for example in an
  \texttt{AUTHORS} file, and to use the aforementioned taxonomy and
  scale.
\end{proposal}

As an illustration of this recommendation, the rich metadata collected
by HAL in the deposit process are sent to SWH using the now standard
CodeMeta schema~\cite{CodeMeta}, and will be soon extended with the
taxonomy of Section~\ref{ssec:taxonomy}

\subsection{Distinguish citation from reference}

We have extensively covered the best practices for assessing and
attributing software artefacts: they are essential for giving
qualified \emph{academic credit} to the people that contribute to
them, and are key prerequisites for creating \emph{citations} for
software. This complex undertaking requires significant human
intervention, and proper processes and tools.

The overall problem of reproducible research is quite different: while
there are examples of rather comprehensive solutions in very
specialised domains, it seems very difficult to find a unique solution
general enough to cover all the use cases. An example of domain
specific solution is provided by the IPOL journal (Image Processing On
Line, \url{https://www.ipol.im/}, an Open Science journal dedicated to
image processing): Each article describes an algorithm and contains
its source code, with an online demonstration facility and an archive
of experiments.

We believe that the three building blocks described in
Section~\ref{ssec:reprod} (Software Heritage, NiX/GUIX, curated
connections between SWH and HAL) will allow to provide precise
references (as illustrated in the end of Section~\ref{sec:complex}) to
both specific software excerpt, context, and environment and to
permanently bind them with research articles.

\begin{proposal}{Distinguish citation from reference}
  \label{rec:cite}
  It is essential to distinguish citations to projects or results from
  exact references to software and their environment, and we believe
  that both should be used in articles. We also strongly encourage
  the use of tools like GUIX and Software Heritage to build such
  perennial references.
\end{proposal}

Although neither a consensus nor a standard exists yet on how to use references in articles,
we are currently working on proposing concrete guidelines and adding support in Software Heritage to easily provide the corresponding \LaTeX~snippets.

\section{Conclusion}
\label{sec:conclusion}

It this article we presented for the first time the internal processes
and efforts in place at Inria for assessing, attributing, and referencing research
software. They play an essential role for the careers of individual
Inria researchers and engineers, the evaluation of whole research
teams, the technology transfer activities and incentive policies, and
the visibility of research teams.

These processes have to cope with the great complexity and variability
of research software, in terms of the nature of its relating
activities and practices, roles of its contributing actors, and
diversity of lifespans.

\subsection*{Recommendations}

Based on our experience over several decades, we have distilled the
important lessons learned and are happy to provide a set of
recommendations that can be summarized as follows:

\begin{description}[leftmargin=\descmargin]

\item[{\bf Recognise the diversity of contributor roles}]\mbox{}\\ The
  taxonomy of contributors described in Section~\ref{ssec:taxonomy}
  has been extensively tested internally at Inria. We recommend that
  it be \emph{incorporated in the CodeMeta standard}, and all the
  platforms and tools that support software attribution and
  citation. In the meanwhile, researchers can adopt it right away in
  the metadata they incorporate in their own source code.

\item[{\bf Keep the human in the loop}]\mbox{}\\ To obtain quality
  metadata, as seen in Section~\ref{ssec:humanintheloop}, it is
  essential to have humans in the loop. We strongly advise against the
  unsupervised use of automated tools to create such metadata. While these
  automated tools can save a lot of time, we recommend instead the
  implementation of a metadata \emph{curation and moderation mechanism} 
	in all tools and platforms that are involved
  in the creation of metadata for research software, like Zenodo or
  FigShare. We also recommend that research institutions and academia
  in general \emph{rely on human experts} to assess the qualitative
  contributions of research software, and refrain from adopting as
  evaluation criteria automated metrics that are easily biased.

\item[{\bf Distinguish citation from reference}]\mbox{}\\ As explained
  in Section~\ref{sec:complex}, \emph{citations}, used to provide
  credit to contributors, are conceptually different from
  \emph{references} designed to support reproducibility. While the
  latter can be largely automated using platforms like Software
  Heritage and tools like GUIX, the former require careful human
  curation. Research articles will then be able to provide both
  software citations and software references, and we are currently
  working on concrete guidelines that we will make publicly available.
\end{description}




\end{document}